\documentclass[12pt]{article}
\usepackage{geometry,amsmath,amssymb,enumerate,epsfig,graphicx,hyperref}
\newcommand{\Omicron}{\mathrm{O}}
\newcommand{\assign}{:=}
\newcommand{\mathe}{\mathrm{e}}
\newcommand{\nocomma}{}
\newcommand{\tmem}[1]{{\em #1\/}}
\newcommand{\tmop}[1]{\ensuremath{\operatorname{#1}}}
\newcommand{\tmtextbf}[1]{{\bfseries{#1}}}

\begin{document}

\title{Finite size scaling and triviality of $\phi^4$ theory on an
antiperiodic torus}

\author{
Matthijs Hogervorst$^{\,a,b}$ and Ulli Wolff$^{\,a,}$\thanks{
e-mail: uwolff@physik.hu-berlin.de}\\[2mm]
{\small $^a$ ~\sl Institut f\"ur Physik, Humboldt Universit\"at,
Newtonstr.~15}\\[-1mm]
{\small \sl 12489~Berlin, Germany}\\
{\small $^b$ ~\sl D\'epartement de Physique, Ecole
Normale Sup\'erieure, 24, rue Lhomond}\\[-1mm]
{\small \sl 75005 Paris, France}
}
\date{}
\maketitle

\begin{abstract}
  Worm methods to simulate the Ising model in the Aizenman random current
  representation including a low noise estimator for the connected four point
  function are extended to allow for antiperiodic boundary conditions. In this
  setup several finite size renormalization schemes are formulated and studied
  with regard to the triviality of $\phi^4$ theory in four dimensions. With
  antiperiodicity eliminating the zero momentum Fourier mode a closer
  agreement with perturbation theory is found compared to the periodic torus.
\end{abstract}
\begin{flushright} HU-EP-11-40 \end{flushright}
\vspace{-0.8cm}
\begin{flushright} SFB/CCP-11-47 \end{flushright}
\thispagestyle{empty}
\newpage

\section{Introduction}

It is generally believed that the quantum field theory of self-coupled scalar
fields in four dimensions is trivial. Then all effects of interaction terms go
away in the true continuum limit. This is certainly true in perturbation
theory due to the universal positivity of the perturbative $\beta$-function at
small coupling (see section \ref{betasst} for more details). In this framework
one also understands that there is a range of values for the mass and
self-coupling such that in the effective theory at finite cutoff there can be
both substantial interaction and only tiny cutoff effects in physical
quantities referring to energies much smaller than the cutoff. This is the
reason why the appearance of a scalar Higgs field in the standard model may
not really be in conflict with triviality. Turned around, triviality even
implies order of magnitude bounds for parameters in the Higgs sector.

There also is the logical possibility that scalar theories could have sectors
for which perturbation theory is simply irrelevant. In a series of papers that
began with {\cite{Luscher:1987ay}} L\"uscher and Weisz have provided a lot of
evidence that in the standard lattice formulation of the theory there is no
such nonperturbative sector for low energy observables, which was also
confirmed early on by Monte Carlo simulation {\cite{Montvay:1988uh}}. In most
numerical investigations the four dimensional Ising model is simulated which
arises as the infinite bare coupling limit of $\phi^4$ theory. While it is
plausible that this is the `least trivial' case the main reason for this
choice is the availability of particularly efficient simulation methods like
the cluster algorithm employed in {\cite{Montvay:1988uh}}.

Triggered by {\cite{prokofev2001wacci}} a further boost in efficiency could be
realized for Ising simulations. The simulation strategy of Prokof'ev and
Svistunov consists of sampling certain strong coupling graphs to arbitrary
order instead of spin configurations. For arbitrary finite systems the
expansion converges and the two representations are equivalent. The authors
propose a simple update scheme for the graph ensembles which does practically
not suffer from critical slowing down. This was elaborated in
{\cite{Wolff:2008km}} by showing that there are in addition dramatically
improved estimators available for two-point correlations. A further step was
made in {\cite{Wolff:2009ke}} where it was noted that the simulated strong
coupling form coincides with the random current representation constructed by
Aizenman {\cite{Aizenman:1982ze}}. He has proved correlation identities that
allowed him to rigorously prove triviality for the Ising model with more than
four dimensions (excluding however $D = 4$). A decisive achievement was a
combinatoric construction of an estimator for the {\tmem{connected}}
four-point function that analytically subtracts the disconnected part and then
could be bounded sharply enough. In the numerical approach at tracing the
renormalized interaction strength {\cite{Montvay:1988uh}} the corresponding
cancellation had to be performed numerically which leads to much reduced
precision. In {\cite{Wolff:2009ke}} Aizenman's identity could be integrated
into a numerical framework by simulating two replica and identifying the
required percolation clusters. It was demonstrated to lead to precise
estimates of renormalized couplings with only moderate computing power in
dimensions $D = 3, 4, 5$.

The question of triviality is an issue of ultraviolet renormalization. In
numerical simulations at $D = 4$ one is limited to ratios $L / a \lesssim
\Omicron \left( 100 \right)$ at present, where $a$ stands for the lattice
spacing and $L$ is the system size. To investigate ultraviolet behavior close
to the continuum limit a clever use of resources is achieved if one uses $L$
itself as the scale to formulate renormalization conditions, i.e. to use a
finite volume scheme as one does with the Schr\"odinger functional in QCD.
Such a strategy - on a simple periodic torus -- was in fact followed in
{\cite{Wolff:2009ke}} by fixing $z = m L$ where $m$ is the renormalized mass
in the so-called second moment definition. It turned out that the first series
of experiments at $z = 2$ led to a borderline agreement with perturbation
theory: The leading order (one loop) described well the cutoff evolution of
the renormalized coupling. The inclusion of two further orders of the
asymptotic expansion however led away from the data. This situation was
further investigated in {\cite{Weisz:2010xx}} with the following results: At
$z = 4$, closer to the thermodynamic limit, perturbation theory works as
expected yielding a successively improving precise description of the
numerical data for the available three orders. For smaller systems it was
argued that the single constant zero momentum mode is responsible for the bad
`convergence' of perturbation theory. This was supported by finding that an
improved expansion, where this mode was treated exactly, yields a much better
agreement with the data also at $z = 2$ and even for $z = 1$.

In the present paper we have eliminated the constant mode by introducing
antiperiodic boundary conditions in one or several directions of the four
dimensional torus. Our new results here consist of generalizing the strong
coupling/random current formulation to these cases including a proof of
Aizenman's identities in section 2. In section 3 we define several finite size
renormalization schemes and report values of the corresponding three loop
$\beta$-function coefficients. In section 4 numerical results for the
antiperiodic case are reported followed by some conclusions. Some details of
the proof of the Aizenman formula and of the perturbative calculations are
deferred to appendices.

\section{Random current form of the Ising model with antiperiodic boundary
conditions}

\subsection{Partition functions with charge insertions}

We here generalize the content of {\cite{Wolff:2009ke}} while we at the same
time slightly change the notation. We now start from the partition function
\begin{equation}
  Z [q] = 2^{- L^4} \sum_s \mathe^{2 \kappa \sum_{l = \langle x y \rangle} z
  (l) s (x) s (y)} \prod_x (s (x))^{q (x)} . \label{Zx1n}
\end{equation}
We independently sum over $s (x) = \pm 1$ on all sites of a four{\footnote{We
discuss $D = 4$ here, but the generalization to other $D$ is trivial.}}
dimensional hypertorus. Here and below all sites are understood to have
integer coordinates with $0 \leqslant x_{\mu} < L$ in all directions $\mu = 0,
1, 2, 3$. On the nearest neighbor links $l$ a Z(2) background gauge
field{\footnote{We prefer $z (l)$ over the usual notation $z (x, \mu)$ here
because links are unoriented.}} $z (l) = \pm 1$ enters and $q (x)$ are fixed
integer local charges whose values enter only modulo 2. For a collection of
sites $x^{\left( 1 \right)}, x^{\left( 2 \right)}, \ldots, x^{\left( n
\right)}$ we may define
\begin{equation}
  q_{12 \ldots n} (x) = \sum_{i = 1}^n \delta_{x, x^{\left( i \right)}}
  \hspace{1em} (\tmop{mod} 2)
\end{equation}
and then a two point function is for example given by
\begin{equation}
  \langle s (x^{\left( 1 \right)}) s (x^{\left( 2 \right)}) \rangle = \frac{Z
  [q_{12}]}{Z [0]} .
\end{equation}
The sole reason for having the background gauge field here is to allow for
antiperiodic boundary conditions. We shall set
\begin{equation}
  z_{\varepsilon} (l = \langle x y \rangle) = \left\{ \begin{array}{lll}
    (- 1)^{\varepsilon_{\mu}} & \tmop{if} & \{x_{\mu}, y_{\mu} \} = \{0, L -
    1\}\\
    + 1 & \tmop{else} & 
  \end{array} \right. .
\end{equation}
Here $\varepsilon_{\mu}$ is a 4-vector with zeros for the periodic and ones
for the antiperiodic directions. Thus there is a minus sign on those links
that `close around the torus' in antiperiodic directions. This is equivalent
to (partially) antiperiodic boundary conditions for the spin field.

We next expand (\ref{Zx1n}) in $\kappa$ by introducing an integer link field
$k (l)$ summed over values $0, 1, 2, \ldots, \infty$ (independently on each
link) and, after summing over the original $s (x)$, we arrive at
\begin{equation}
  Z [q] = \sum_k w [k] \Phi_{\varepsilon} [k] \delta_{\partial k, q}
\end{equation}
with
\begin{equation}
  w [k] = \prod_l \frac{(2 \kappa)^{k (l)}}{k (l) !} . \label{wdef}
\end{equation}
Here the divergence of $k$
\begin{equation}
  \partial k (x) = \sum_{l, \partial l \ni x} k (l) \hspace{1em} (\tmop{mod}
  2)
\end{equation}
is a site field where we add the $k (l)$ of the eight links surrounding $x$.
It is {\tmem{locally}} constrained to equal (modulo 2) the source $q (x)$. The
sign
\begin{equation}
  \Phi_{\varepsilon} [k] = \prod_l [z_{\varepsilon} (l)]^{k (l)} . \label{Phi}
\end{equation}
is the Z(2) winding number of the $k$ field with respect to the antiperiodic
directions coded into $\varepsilon$.

\subsection{Connected four point function}

If we introduce{\footnote{There is a trivial error in eq. (9) of
{\cite{Wolff:2009ke}}.}}
\begin{equation}
  Z_c (q_{1234}) = Z [q_{1234}] Z [0] - Z [q_{12}] Z [q_{34}] - Z [q_{13}] Z
  [q_{24}] - Z [q_{14}] Z [q_{23}]
\end{equation}
then the connected four point function is given by
\begin{equation}
  \langle s (x^{\left( 1 \right)}) s (x^{\left( 2 \right)}) s (x^{\left( 3
  \right)}) s (x^{\left( 4 \right)}) \rangle_c = \frac{Z_c [q_{1234}]}{Z
  [0]^2} .
\end{equation}
With the help of the results of appendix \ref{appa} this can be written as
\begin{equation}
  Z_c (q_{1234}) = - 2 \sum_{k, k'} w [k] w [k'] \Phi_{\varepsilon} [k]
  \Phi_{\varepsilon} [k'] \delta_{\partial k, q_{12}} \delta_{\partial k',
  q_{34}} \mathcal{X}_{13}, \label{Aizlemma}
\end{equation}
where we have inserted
\begin{equation}
  \Phi_{\varepsilon} [k + k'] = \Phi_{\varepsilon} [k] \Phi_{\varepsilon} [k']
\end{equation}
for the arbitrary function $F [k + k']$. The constraint $\mathcal{X}_{13}
\equiv \mathcal{X} (x^{\left( 1 \right)}, x^{\left( 3 \right)} ; k + k') \in
\{0, 1\}$ is one if and only if $x^{\left( 1 \right)}$ and $x^{\left( 3
\right)}$ are in the same percolation cluster with respect to bonds which are
active on links where $k + k'$ does not vanish.

We now introduce an ensemble of two independent (factorizing) replica
\begin{equation}
  \mathcal{Z}= \sum_{u, v, k} \sum_{u', v', k'} w [k] w [k'] \delta_{\partial
  k, q_{u v}} \delta_{\partial k', q_{u' v'}} \label{Zens}
\end{equation}
with
\begin{equation}
  q_{u v} (x) = \delta_{x, u} + \delta_{x, v}
\end{equation}
and expectation values $\langle \langle \ldots \rangle \rangle$ in this
ensemble are defined in the obvious way. In particular, the two point function
now reads
\begin{equation}
  \left\langle s (x) s (y) \right\rangle = \frac{\left\langle \left\langle
  \Phi_{\varepsilon} [k] \delta_{x, u} \delta_{y, v} \right\rangle
  \right\rangle}{L^{- 4} \left\langle \left\langle \Phi_{\varepsilon} [k]
  \delta_{u, v} \right\rangle \right\rangle} \label{twopoint}
\end{equation}
with the sign being part of the observables. This may of course be symmetrized
over the two replica. For the connected four point function we get
\begin{equation}
  \langle s (x^{\left( 1 \right)}) \cdots s (x^{\left( 4 \right)}) \rangle_c =
  - 2 \frac{\left\langle \left\langle \right. \right. \Phi_{\varepsilon} [k]
  \Phi_{\varepsilon} [k'] \delta_{x^{\left( 1 \right)}, u} \delta_{x^{\left( 2
  \right)}, v} \delta_{x^{\left( 3 \right)}, u'} \delta_{x^{\left( 4 \right)},
  v'} \mathcal{X} \left( u, u' ; k + k' \right) \left\rangle
  \right\rangle}{L^{- 8} \left\langle \left\langle \Phi_{\varepsilon} [k]
  \Phi_{\varepsilon} [k'] \delta_{u, v} \delta_{u', v'} \right\rangle
  \right\rangle} \label{con4}
\end{equation}
Note that for nonzero contributions, all four charges at $u, v, u', v'$ are in
the same percolation cluster.

\section{Renormalized mass and coupling}

\subsection{Families of finite volume renormalization schemes}

Employing (partially) antiperiodic boundary conditions we now define a choice
of possible renormalization conditions that use the finite system size as
renormalization scale. The sets $\mathcal{B}_{\varepsilon}$ of admissible
momenta depend on $\varepsilon$ and are given by
\begin{equation}
  \phi (p) = \sum_x \mathe^{- i p x} s (x), \hspace{1em} p \in
  \mathcal{B}_{\varepsilon} = \left\{ p_{\mu} = (n_{\mu} + \varepsilon_{\mu} /
  2) \times 2 \pi / L, 0 \leqslant n_{\mu} < L \right\} . \label{Brill}
\end{equation}
The zero momentum mode, which has led to a bad convergence of renormalized
perturbation theory {\cite{Weisz:2010xx}} on small tori, does not occur
anymore for $\varepsilon \neq (0, 0, 0, 0)$.

We use an index `$p$' for the periodic case ($\varepsilon_{\mu}^{\left( p
\right)} \equiv 0$), `$A$' for fully antiperiodic ($\varepsilon_{\mu}^{\left(
A \right)} \equiv 1$) and `$a$' for one antiperiodic direction
($\varepsilon_{\mu}^{\left( a \right)} \equiv \delta_{\mu, 0}$) and then also
write $\mathcal{B}_s =\mathcal{B}_{\varepsilon^{\left( s \right)}}$. For each
case $s \in \{p, a, A\}$ we single out two small admissible momenta $p_s$ and
$p_s'$ with distinct $\hat{p}^2_s < \hat{p}_s'^2$ with
\begin{equation}
  \hat{p}^2 = 4 \sum_{\mu} \sin^2 (p_{\mu} / 2),
\end{equation}
namely
\begin{equation}
  p_s = \frac{\pi}{L} \varepsilon^{\left( s \right)}, \hspace{1em} p_s' = p_s
  + \left( 0, 0, 0, 2 \pi / L \right).
\end{equation}
Renormalized masses $m_s$ are defined by solving for $m_s$ in the universal
ratios
\begin{equation}
  R_s = \frac{\langle | \phi (p_s') |^2 \rangle}{\langle | \phi (p_s) |^2
  \rangle} = \frac{\hat{p}_s^2 + m_s^2}{\hat{p}_s'^2 + m_s^2} \hspace{1em}
  \Rightarrow z_s = m_s L. \label{mdef}
\end{equation}
We note that in the symmetric phase massive scaling regions defined by $0 <
m_s^2 \ll 1$ we must adjust $R_p \gtrsim 0$, $R_a \gtrsim 1 / 5$ and $R_A
\gtrsim 1 / 3$ (up to O($L^{- 2}$)).

To define corresponding renormalized coupling constants $g_s$ we employ the
smaller of the two momenta and form another ratio
\begin{equation}
  g_s = - \frac{\langle | \phi (p_s) |^4 \rangle_c}{\langle | \phi (p_s) |^2
  \rangle^2}  (z^2_s + L^2 \hat{p}_s^2)^2 .
\end{equation}
Note the {\tmem{connected}} four point function in the numerator. The case
$m_p, g_p$ coincides with the scheme studied in {\cite{Wolff:2009ke}}. Each
definition of $g_s$ has the following properties:
\begin{itemize}
  \item It derives from universal renormalized ratios of correlations with
  wavefunction renormalization factors canceling.
  
  \item If applied to $\phi^4$ theory away from the Ising limit, it coincides
  with the standard bare coupling at tree level of perturbation theory.
  
  \item For $z_s \rightarrow \infty$ boundary conditions become irrelevant and
  all $g_s$ coincide with the usual coupling defined by vertex functions at
  zero momentum, and $m_s$ approaches the infinite volume mass defined at zero
  momentum. We hence make contact with the scheme of {\cite{Luscher:1987ay}}.
\end{itemize}
We may now parameterize the renormalized theory by fixing a mass and a
coupling constant. In the most general case we may even choose different
boundary conditions $s, t \in \left\{ p, a, A \right\}$ for this purpose and
approach the finite volume continuum limit $L \equiv L / a \rightarrow \infty$
for fixed $z_t$ and $g_s$.

\subsection{Beta functions}\label{betasst}

For each set of normalization conditions we consider the Callan-Symanzik
evolution equation with the cutoff $L \equiv L / a$
\begin{equation}
  L \frac{\partial g_s}{\partial L} \left|_{z_t = z} \right. = - \beta_{s, t,
  z} (g_s) .
\end{equation}
As is customary in the $\phi^4$ literature {\cite{Luscher:1987ay}} we take the
derivative on the left hand side at fixed {\tmem{bare}} coupling. Terms of
order $L^{- 2}$ are neglected and therefore $\beta_{s, t, z}$ is a function of
$g_s$ only. In the perturbative expansion
\begin{equation}
  \beta_{s, t, z} (g) = \sum_{l \geqslant 1} b_{s, t, z}^{(l)} g^{l + 1}
\end{equation}
the first two coefficients are scheme independent,
\begin{equation}
  b_{s, t, z}^{(1)} = \frac{3}{(4 \pi)^2}, \hspace{1em} b_{s, t, z}^{(2)} =
  \frac{17 / 3}{(4 \pi)^4},
\end{equation}
while the third coefficient is known in the infinite volume
{\cite{Luscher:1987ay}}
\begin{equation}
  b_{s, t, \infty}^{(3)} = \frac{26.908403}{(4 \pi)^6} .
\end{equation}

Knowing the two loop relation between couplings in two schemes allows to
compute the {\tmem{difference}} between their respective coefficients
$\beta_{s, t, z}^{(3)}$. All necessary formulas are found in section 3.2 of
{\cite{Weisz:2010xx}}. In the same paper, from Table~1, values for $b_{p, p,
z}^{(3)}$ (the scheme used there) for many different $z$ are given by relating
them in steps to large, effectively infinite, $z$. We use these values now and
connect to them the schemes relevant in this paper. By working out the
necessary Feynman diagram sums (see appendix \ref{appB1} for details) up to $L
= 100$ we have obtained Table~\ref{db3}.

\begin{table}[htb]
\begin{center}
  \begin{tabular}{|r|r@{.}l|r@{.}l|}
    \hline
    $s, t, z$ & \multicolumn{2}{c|}{$(b_{s, t, z}^{(3)} - b_{p, p, z}^{(3)}) \times (4 \pi)^6$} &
    \multicolumn{2}{c|}{$b_{s, t, z}^{(3)} \times (4 \pi)^6$}\\
    \hline
    $p, a, 2$ & 237&2805 & 646&3126\\
    \hline
    $a, a, 2$ & $-374$&8514 & 34&1807\\
    \hline
    $A, a, 2$ & $-377$&5345 & 31&4976\\
    \hline
    $p, a, 3$ & 4&65458 & 40&9003\\
    \hline
    $a, a, 3$ & $-7$&07988 & 29&1658\\
    \hline
    $A, a, 3$ & $-6$&20041 & 30&0453\\
    \hline
  \end{tabular}
  \caption{Three-loop coefficients of the $\beta$-functions for the boundary
  conditions relevant in this study, i.e. mass $z_a = m_a L = 2, 3$ and
  couplings $g_p, g_a, g_A$. \label{db3}}
\end{center}
\end{table}

\subsection{Estimators}

The correlations entering into $m_s$ and $g_s$ are now translated into
expectation values in the ensemble (\ref{Zens}). It is not difficult to find
for the ratio in (\ref{mdef})
\begin{equation}
  R_s = \frac{\left\langle \left\langle \Phi_s [k] f_s' (u - v) \right\rangle
  \right\rangle}{\left\langle \left\langle \Phi_s [k] f_s (u - v)
  \right\rangle \right\rangle}, \hspace{1em} \Phi_s \equiv
  \Phi_{\varepsilon^{\left( s \right)}}
\end{equation}
with
\begin{equation}
  f_s (x) = \prod_{\mu} \cos (x_{\mu} p_{s, \mu}), \hspace{1em} f_s' (x) =
  \prod_{\mu} \cos (x_{\mu} p_{s, \mu}'),
\end{equation}
The invariance with respect to separate reflections of each direction has been
used to factorize the Fourier exponentials into cos factors{\footnote{Omitted
parts with sin factors would average to zero but still contribute noise. In
principle, if it is {\tmem{anticorrelated}} with the signal, this could lower
the error, but this is unlikely.}}. Where $\varepsilon^{\left( a \right)},
p_a, p_a'$ and $p_A'$ treat the four directions differently we average over
all possible ways of putting the anisotropies in our observables. In addition
we combine the two replica in the error analysis as discussed in
{\cite{Wolff:2003sm}}.

For the coupling, a possible estimator is given by
\begin{equation}
  g_s = 2 (z^2_s + L^2 \hat{p}_s^2)^2 \mathcal{X}_s
\end{equation}
with
\begin{equation}
  \mathcal{X}_s = \frac{\left\langle \left\langle \Phi_s [k] \Phi_s [k'] f_s
  (u + u' - v - v')\mathcal{X}(u, u' ; k + k') \right\rangle
  \right\rangle}{\left\langle \left\langle \Phi_s [k] \Phi_s [k'] f_s (u - v)
  f_s (u' - v') \right\rangle \right\rangle} \label{Xest}
\end{equation}
This is however a special choice. Since the left hand side of (\ref{con4}) is
symmetric in its arguments, one could also permute the arguments in $f_s (u +
u' - v - v')$, inequivalent choices being $f_s (u - u' + v - v')$ and $f_s (u
- u' - v + v')$ in addition. We found it quite profitable in terms of errors
to average over the three possibilities (after verifying that their mean
values are compatible). The denominator factorizes of course in the replica,
and also here other choices are possible{\footnote{The one given looks most
natural and promising.}}.

We note that the factors $L^2 \hat{p}_A^2 \approx 4 L^2 \hat{p}_a^2 \approx 4
\pi^2$ will enhance the values of the actually measured observable quite
significantly for the antiperiodic cases. Correspondingly $\mathcal{X}_{a, A}$
will typically be found much smaller than $\mathcal{X}_p$ due to
cancellations. The same cancellations -- absent in $\mathcal{X}_p$ which has a
non-negative estimator -- will also lead to lower achievable precision for the
antiperiodic cases, in particular for `$A$'. The fluctuating sign included in
our observables is however not of the kind that leads to an exponential signal
to noise problem on large lattices. The signs are `coherently' related to the
winding around the torus and to Fourier modes with wave numbers of order $1 /
L$ and do not combine nearly independent signs from many small subvolumes. The
latter is typical when `sign problems' render numerical estimates impossible.

\subsection{Couplings from partition function ratios}

In our finite volume simulations we also obtain information about ratios of
partition functions with differing boundary conditions, for example
\begin{equation}
  \frac{Z_a}{Z_A} = \frac{\langle \langle \Phi_a \delta_{u, v} \rangle
  \rangle}{\langle \langle \Phi_A \delta_{u, v} \rangle \rangle} .
\end{equation}
From the form of these estimators where signs are averaged it is trivial that
$Z_a \leqslant Z_p$ and $Z_A \leqslant Z_p$ holds (in the Ising limit!), and
in addition $Z_A \leqslant Z_a$ is plausible and indeed found numerically in
all cases. In the Gaussian limit discussed below the same ordering holds.

In perturbation theory we may expand the differences in free energy
\begin{equation}
  \ln (Z_t / Z_s) = f_0^{t, s} (z_t, L / a) + f_1^{t, s} (z_t, L / a) g_0 +
  \Omicron (g_0^2) \label{ZZPT}
\end{equation}
and some values of $f_0, f_1$ are listed in appendix \ref{appPT}.

We may hence define further conventionally normalized coupling constants by
\begin{equation}
  h_{t / s} = \frac{\ln (Z_t / Z_s) - f_0^{t, s}}{f_1^{t, s}} .
\end{equation}
They measure the response of the free energy to a change of boundary
conditions and are expected to be physical quantities in the continuum limit
and thus legitimate renormalized couplings in a finite volume. Note that the
renormalized mass $z_t$ is chosen on the right hand side of (\ref{ZZPT}). To
the order cited we may also just replace $g_0$ by any renormalized coupling.
Thus $f_{0, 1}^{t, s}$ refer to relations between renormalized quantities and
are expected and indeed found to reach finite limits as $L / a \rightarrow
\infty$. A numerical disadvantage of $h_{t / s}$ is the required subtraction
of the tree level part. After this cancellation, the precision of $h_{a / A}$
in our present simulation is too low for a meaningful study of its evolution.
Phrased differently, in our simulations $Z_a / Z_A$, which is typically per
mil accurate, is given within errors by its free field value, another result
that is consistent with triviality.

\section{Numerical results}

We have developed a serial C-code to sample the graphs of the ensemble
contributing in (\ref{Zens}). Details are very similar to those given in
{\cite{Wolff:2009ke}} except that we have this time used a Metropolis rather
than a heatbath step to move the worm-ends. An iteration consists of $L^4$
(attempted) worm moves for each of the two replica with about 64 percolation
processes interspersed to compute $\mathcal{X}$. The signs $\Phi_s$ are
`updated' as the worms move and are hence available at any time to
continuously accumulate observables during the updates. An iteration requires
a computational effort proportional to $L^4$ that is roughly comparable to a
sweep in a local update scheme. For each data point (value of $L$ and
$\kappa$) we executed $10^6$ iterations (after equilibration) where we always
stored the blocked measurements from 10 successive iterations. Thus we had to
analyze time-series of length $10^5$ and in these units we found at most
integrated autocorrelation times of unity and for many observables the absence
of any relevant correlations. Equilibration under these circumstances has been
unproblematic.

We have exploited some trivial parallelization to generate our data by running
between 4 and 64 copies of the system to produce the total statistics. The
runs took place on dual-quad-core X2270 PCs. Each of the $L = 64$ runs took
about 3000 core-hours or about 2 days with 64 cores and the smaller lattices
follow by scaling proportional to $L^4$.

\begin{table}[htb]
\begin{center}
  \begin{tabular}{|l|l|l|c|l|l|l|}
    \hline
    $L$ & \multicolumn{1}{c|}{$2 \kappa$} & \multicolumn{1}{c|}{$z_a$} & 
    $-\frac{\kappa}{L^2} \frac{\partial z_a}{\partial \kappa}$ & 
    \multicolumn{1}{c|}{$z_p$} & \multicolumn{1}{c|}{$z_a$} & 
    \multicolumn{1}{c|}{$z_A$}\\
    \hline
    \phantom{1}8 & .1475570 & 1.9988(68) & 1.0804(52) & $2.2587 \left( 36 \right)$ & $2.0$ &
    $1.781 \left( 24 \right)$\\
    \hline
    10 & .1482830 & 2.0003(66) & 1.0440(50) & 2.2404(35) & 2.0 & 1.835(21)\\
    \hline
    12 & .1486864 & 1.9948(64) & 1.0272(49) & 2.2270(34) & 2.0 & 1.843(20)\\
    \hline
    16 & .1490990 & 2.0012(60) & 0.9739(47) & 2.2077(33) & 2.0 & 1.843(18)\\
    \hline
    22 & .1493687 & 1.9971(57) & 0.9400(45) & 2.1947(31) & 2.0 & 1.886(16)\\
    \hline
    32 & .1495330 & 2.0015(54) & 0.9043(42) & 2.1806(29) & 2.0 & 1.884(14)\\
    \hline
    64 & .1496509 & 1.9993(48) & 0.8470(39) & 2.1611(26) & 2.0 & 1.912(12)\\
    \hline
  \end{tabular}
  \caption{Numerical results for simulations at $z_a = 2$.\label{tab2}}
\end{center}
\end{table}

\begin{table}[htb]
\begin{center}
  \begin{tabular}{|l|l|l|l|l|}
    \hline
    \multicolumn{1}{|c|}{$g_p$} & \multicolumn{1}{c|}{$g_a$} & 
    \multicolumn{1}{c|}{$g_A$} & \multicolumn{1}{c|}{$h_{a / p}$} & 
    \multicolumn{1}{c|}{$h_{a / A}$}\\
    \hline
    16.44(7) & $30.26 \left( 27 \right)$ & $33 \left( 12 \right)$ & 16.58(52)
    & 33.9(6.6)\\
    \hline
    15.14(6) & 27.19(25) & 36(11) & 14.25(50) & 23.1(6.3)\\
    \hline
    14.25(6) & 24.94(24) & 48(11) & 14.10(48) & 31.8(6.1)\\
    \hline
    12.96(6) & 22.45(22) & 19(10) & 13.26(45) & 19.7(5.6)\\
    \hline
    11.89(5) & 20.23(19) & \phantom{1}7(9) & 12.10(42) & 21.9(5.1)\\
    \hline
    10.87(5) & 17.81(17) & 36(8) & 10.99(38) & 16.9(4.6)\\
    \hline
    \phantom{1}9.34(4) & 14.70(13) & 11(8) & 10.09(31) & 11.5(3.7)\\
    \hline
  \end{tabular}
  \caption{Companion to Table~$\ref{tab2}$ with more observables, lines in the
  same order.\label{tab3}}
\end{center}
\end{table}

In Tables~\ref{tab2} and \ref{tab3} we compile results where for a series of
lattices of sizes $L = 8, \ldots, 64$ we have tuned $\kappa$ to values that
produce $z_a \approx 2$ to a very good approximation. For all our observables
we have measured their $\kappa$-derivatives as connected correlation with $S_k
= \sum_l k \left( l \right)$, for example
\begin{equation}
  \kappa \frac{\partial}{\partial \kappa} \left\langle \left\langle \Phi_s [k]
  f_s' (u - v) \right\rangle \right\rangle = \left\langle \left\langle \Phi_s
  [k] f_s' (u - v) S_k \right\rangle \right\rangle - \left\langle \left\langle
  \Phi_s [k] f_s' (u - v) \rangle \rangle \langle \langle S_k \right\rangle
  \right\rangle .
\end{equation}
The fourth column of Table~\ref{tab2}, relevant for the tuning of
$\kappa$ has for example been
obtained in this way. Moreover
we have implemented a small post-run reweighting to first order in the
$\kappa$-shift to achieve $z_a = 2$ exactly, as already discussed in
{\cite{Wolff:2009ke}}. The overall error estimate for this somewhat involved
function of primary observables was determined following {\cite{Wolff:2003sm}}
with the error of the $\kappa$-derivatives safely neglected for the only small
corrections. Thus the first four columns in Table~\ref{tab2} refer to the
parameters that were actually simulated. The remaining columns as well as the couplings in Table~\ref{tab3}
include the (tiny) corrections and thus refer to $z_a = 2$ as required for
finite size scaling. We see relative errors {\tmem{fall}} with growing $L$. As
we spend about constant computer time per site we experience slightly
{\tmem{negative}} critical slowing down here.

While the estimator (\ref{Xest}) is non-negative in the periodic case this is
not anymore the case for $s = a, A$. In particular for the fully antiperiodic
coupling $g_A$ at $z = 2$ the sign fluctuations are too strong to leave a
useful signal at our statistics.

\begin{table}[htb]
\begin{center}
  \begin{tabular}{|l|l|l|c|l|l|l|}
    \hline
    $L$ & \multicolumn{1}{|c|}{$2 \kappa$} & \multicolumn{1}{|c|}{$z_a$} & 
     $-\frac{\kappa}{L^2} \frac{\partial z_a}{\partial \kappa}$& 
    \multicolumn{1}{|c|}{$z_p$} & \multicolumn{1}{|c|}{$z_a$} & 
    \multicolumn{1}{|c|}{$z_A$}\\
    \hline
    \phantom{1}8 & .1450850 & 2.9968(42) & 0.7938(26) & 3.0815(24) & 3.0 & 2.913(8)\\
    \hline
    10 & .1465910 & 3.0016(39) & 0.7593(25) & 3.0739(22) & 3.0 & 2.919(8)\\
    \hline
    12 & .1474720 & 2.9980(37) & 0.7319(24) & 3.0661(21) & 3.0 & 2.922(7)\\
    \hline
    16 & .1483860 & 2.9983(35) & 0.6946(23) & 3.0615(20) & 3.0 & 2.942(6)\\
    \hline
    22 & .1489732 & 2.9969(34) & 0.6685(23) & 3.0550(18) & 3.0 & 2.944(6)\\
    \hline
    32 & .1493373 & 2.9971(31) & 0.6337(22) & 3.0478(17) & 3.0 & 2.962(5)\\
    \hline
    64 & .1495982 & 2.9988(28) & 0.5898(21) & 3.0417(14) & 3.0 & 2.963(4)\\
    \hline
  \end{tabular}
  \caption{As Table~\ref{tab2} but for $z_a = 3$.\label{tab4}}
\end{center}
\end{table}

\begin{table}[htb]
\begin{center}
  \begin{tabular}{|l|l|l|l|l|}
    \hline
    \multicolumn{1}{|c|}{$g_p$} & \multicolumn{1}{|c|}{$g_a$} & 
    \multicolumn{1}{|c|}{$g_A$} & \multicolumn{1}{|c|}{$h_{a / p}$} & 
    \multicolumn{1}{|c|}{$h_{a / A}$}\\
    \hline
    28.55(10) & 36.86(16) & 42.6(2.5) & 28.8(3.5) & 23(12)\\
    \hline
    25.73(9) & 32.71(14) & 40.7(2.2) & 25.3(3.2) & 22(11)\\
    \hline
    23.84(8) & 29.81(13) & 34.6(2.0) & 26.9(3.0) & 31(11)\\
    \hline
    21.30(7) & 26.38(11) & 28.6(1.8) & 26.4(2.8) &  \phantom{1}7(10)\\
    \hline
    19.13(6) & 23.04(9) & 26.3(1.6) & 15.1(2.6) & 19(9)\\
    \hline
    17.00(5) & 20.30(8) & 22.9(1.4) & 17.1(2.3) & 20(8)\\
    \hline
    14.15(4) & 16.49(6) & 18.9(1.2) & 16.2(1.8) &  \phantom{1}8(7)\\
    \hline
  \end{tabular}
  \caption{Extension of Table~\ref{tab4}.\label{tab5}}
\end{center}
\end{table}

Tables~\ref{tab4} and \ref{tab5} are structured in the same way as 
Tables~\ref{tab2} and \ref{tab3} but refer to physically larger volumes with $z_a =
3.$

Remember that the columns in our tables refer to fixed values of $z_a$ and the
{\tmem{bare}} $\phi^4$ coupling which is infinite in the Ising limit. For
small $a / L$ we expect (almost) universal relations between one pair $\left(
z_t, g_s \right)$ and another one. The values of $z_p$ or $z_A$ following
downwards the columns of Table~\ref{tab2} or \ref{tab4} are expected to
converge slowly at a rate given by the vanishing renormalized coupling (if
triviality holds) to the free field result $z_p = z_A = z_a$ and not to
nontrivial values at rates $L^{- 2}$.

\begin{figure}[htb]
\begin{center}
  \resizebox{0.5\textwidth}{!}{\includegraphics{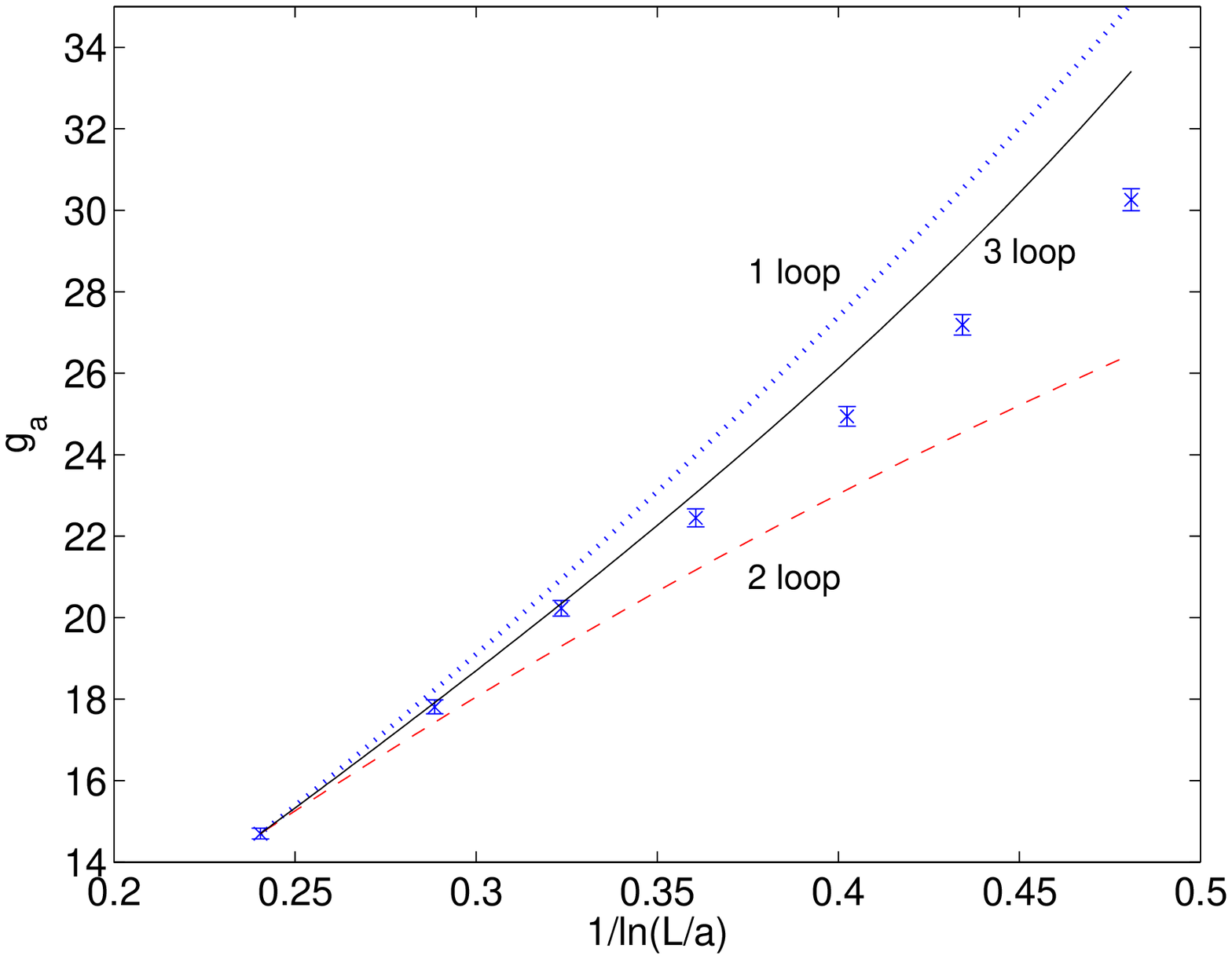}}\resizebox{0.5\textwidth}{!}{\includegraphics{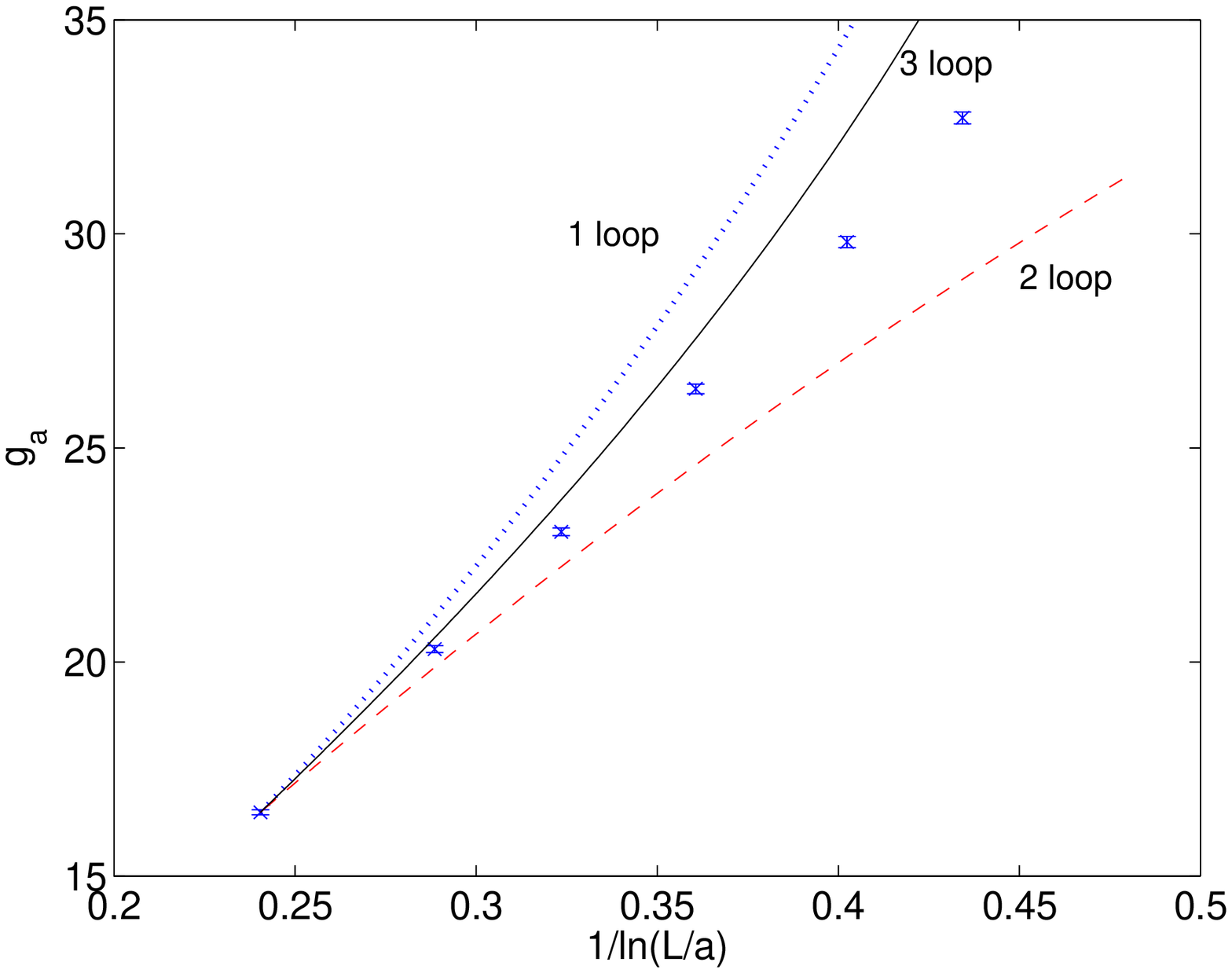}}
  \caption{Cutoff dependence of the coupling $g_a$ at $z_a = 2 \nocomma
  \nocomma$ (left plot) and $z_a = 3$ (right plot).\label{fig1}}
\end{center}
\end{figure}

In Fig.~\ref{fig1} we see the evolutions of $g_a$ at $z_a = 2, 3$. In both
cases we find agreement with the perturbative pattern that systematically
improves with the loop order for the three terms that are available. This is
particularly pronounced if one compares the curves for $z_a = 2$ with the
fully periodic case in {\cite{Wolff:2009ke}}. The expectation that
antiperiodic boundary conditions render all modes perturbative also in a
smaller volume is confirmed.

\begin{figure}[htb]
\begin{center}
  \resizebox{0.5\textwidth}{!}{\includegraphics{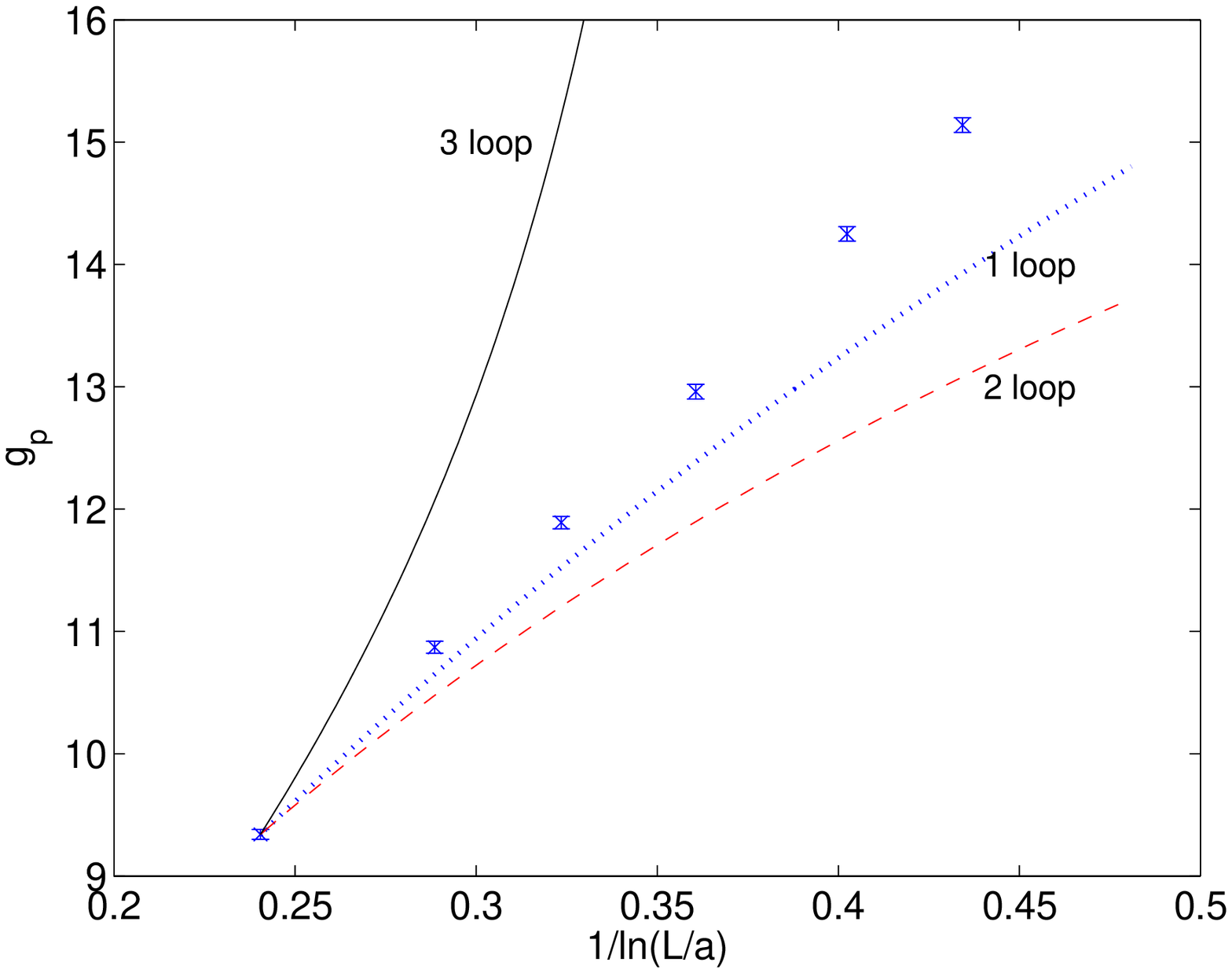}}\resizebox{0.5\textwidth}{!}{\includegraphics{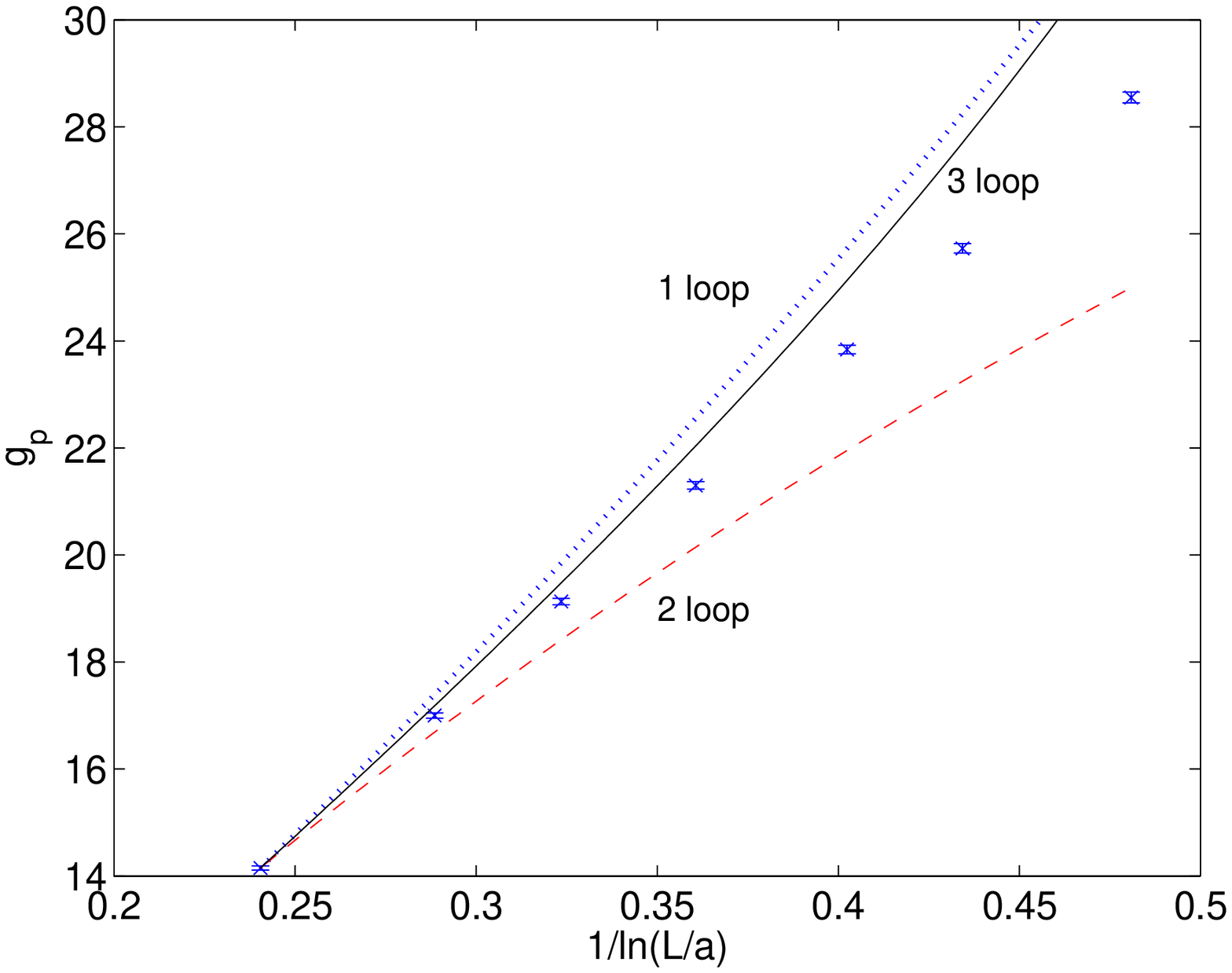}}
  \caption{Cutoff dependence of the coupling $g_p$ at $z_a = 2 \nocomma
  \nocomma$ (left plot) and $z_a = 3$ (right plot).\label{fig2}}
\end{center}
\end{figure}

In Fig.~\ref{fig2} we investigate two schemes in terms the periodic $g_p$
combined with $z_a$. In particular the left plot looks very similar to
{\cite{Wolff:2009ke}}: the antiperiodic mass alone is not sufficient to
eliminate the effects of the constant mode that contributes to $g_p$.

\begin{figure}[htb]
\begin{center}
  \resizebox{0.5\textwidth}{!}{\includegraphics{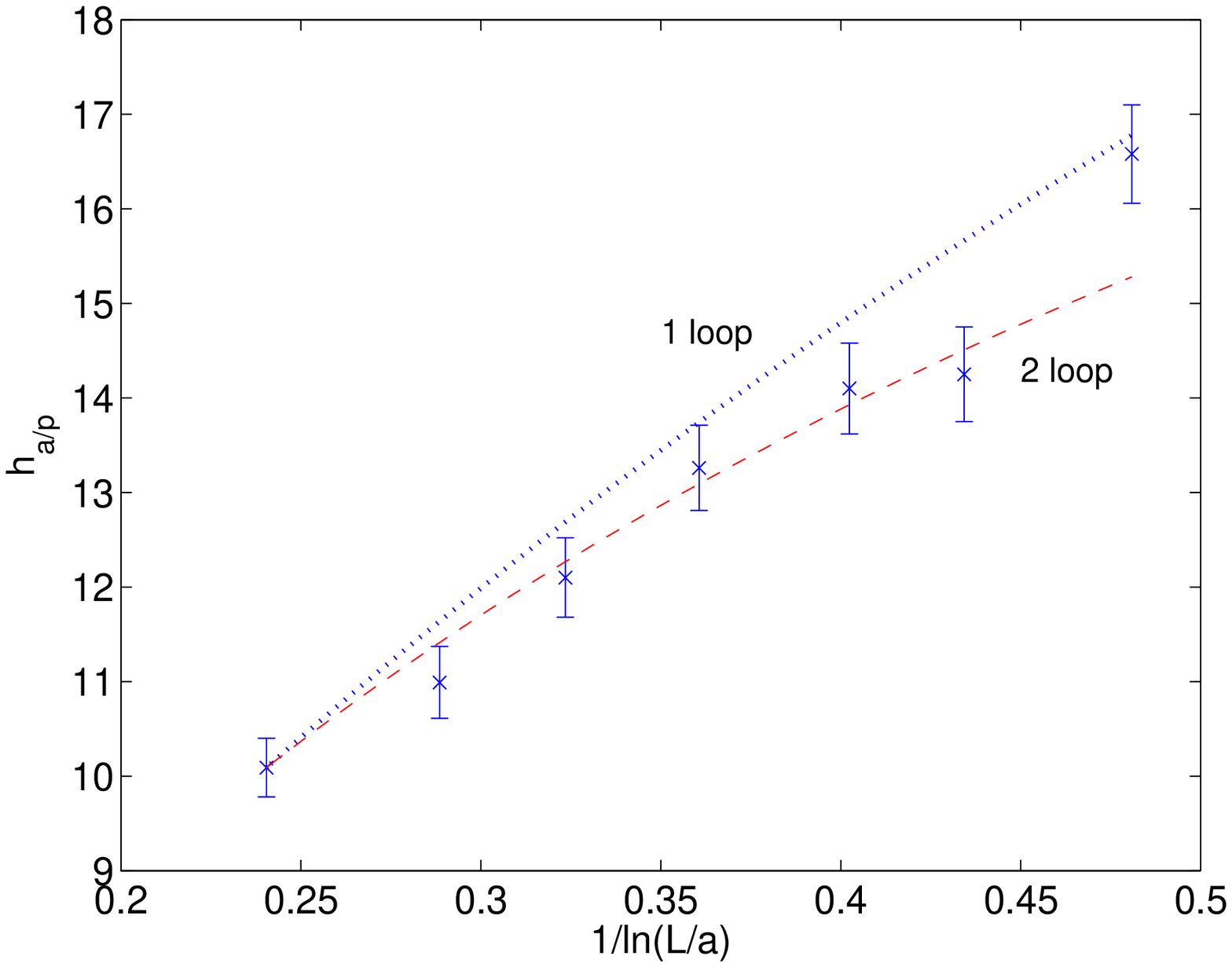}}\resizebox{0.5\textwidth}{!}{\includegraphics{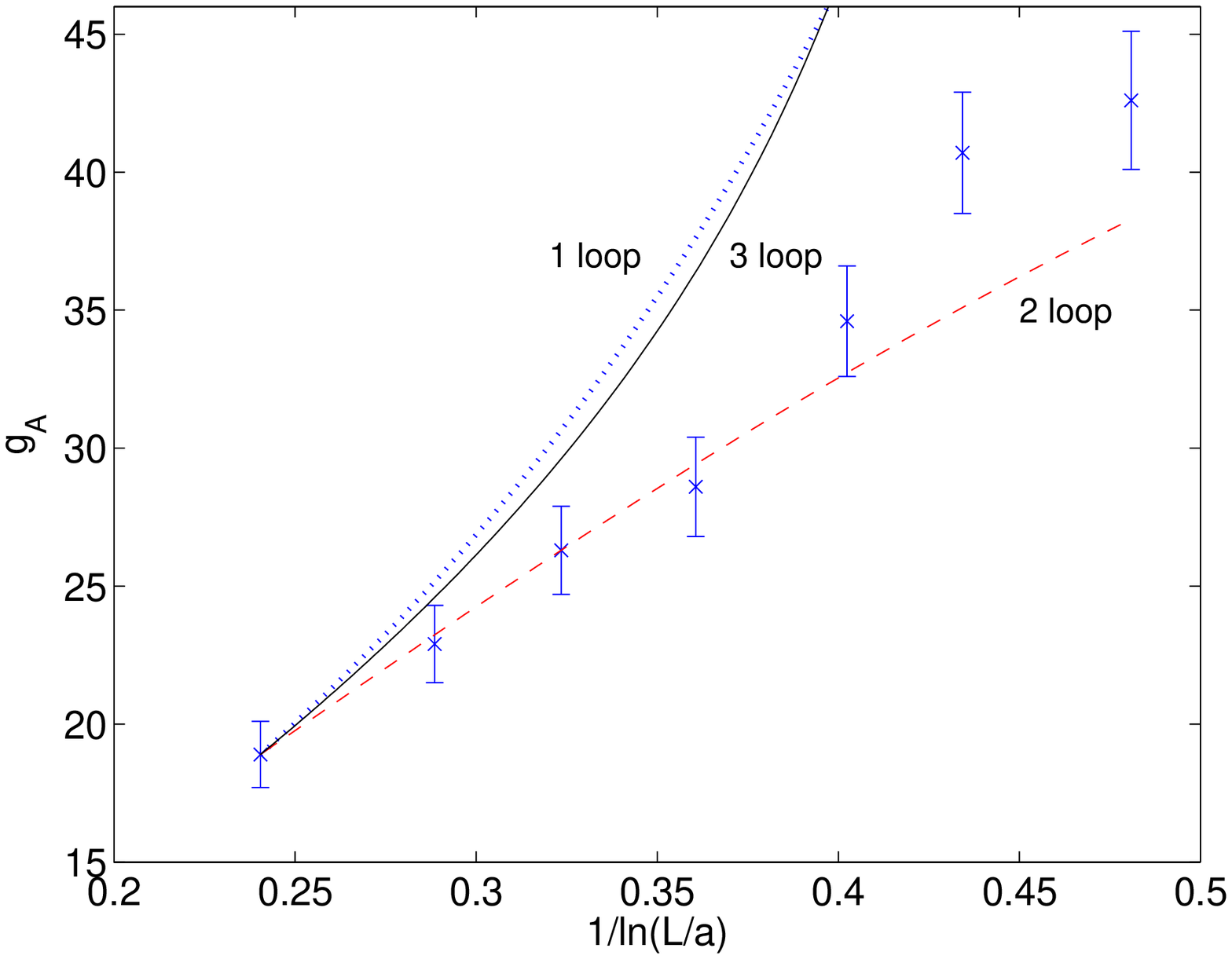}}
  \caption{Cutoff dependence of the coupling $h_{a / p}$ at $z_a = 2 \nocomma
  \nocomma$ (left plot) and for the coupling $g_A$ at $z_a = 3 \nocomma
  \nocomma$ (right plot).\label{fig3}}
\end{center}
\end{figure}

In Fig.~\ref{fig3} we study two couplings which we can compute with lower
but still reasonably significant precision. On the left we see agreement
between $Z_a / Z_p$ at $z_a = 2$ with two loop perturbation theory, where no
three loop term is available. Finally the right plot shows the fully
antiperiodic $g_A$ at $z_a = 3$. Since relatively large momenta contribute
here, the cutoff effects could be larger in this plot than in the previous
ones. While we expect them to be still small for $L = 64, 32$ we cannot really
disentangle them. We refer to the discussion in {\cite{Weisz:2010xx}} that the
perturbative artefacts, that could be specified at 1 and 2 loop order, are not
relevant here.

\section{Conclusions}

In {\cite{Wolff:2009ke}} a Monte Carlo algorithm was presented for simulating
Aizenman's reformulation of the Ising model as a statistical system of random
currents on links {\cite{Aizenman:1982ze}}. This form has two advantages:
(practical) absence of critical slowing down and the availability of a
non-negative estimator for the {\tmem{connected}} four point function, that
enters into the standard definition of renormalized interaction strength,
without having to perform numerical cancellations. In the present paper we
have generalized this from periodic boundary conditions in all four directions
to arbitrary combinations of periodic and antiperiodic directions. Then the
above mentioned estimator starts to fluctuate in sign but still yields good
precision for one antiperiodic direction and system sizes as small as $z = 2$.
With four antiperiodic directions the noise was found to much more degrade the
possible precision. One and the same simulation produces results for all the
boundary conditions considered here. For large $z$ the simulated graphs do not
wind around the torus and the independence of results on the choice of
boundary conditions becomes manifest.

In our previous work for the periodic case we have found, that the decay of
the coupling strength with the UV cutoff (triviality) is not very well
described by perturbation theory given the precision of our new method. This
is a problem, because even very efficient numerical simulations cannot trace
this decay over significant scale-ranges, if it happens only at the expected
logarithmic rate. What it can only do is to confirm the matching with
perturbation theory which can then be trusted all the way to the continuum. In
{\cite{Weisz:2010xx}} we have accumulated some evidence that the constant zero
momentum mode that exists for periodic boundary conditions is the main source
of nonperturbative behavior in small volumes. With at least one antiperiodic
direction the smallest momentum is $\pi / L$ and all modes receive Gaussian
damping independently of the mass term. In Fig.~\ref{fig1} we demonstrate that
perturbation theory indeed works much better now.

In the present simulations we could determine with good precision the change
in free energy caused by differing boundary conditions. Also these quantities
can be related to renormalized couplings and probed numerically. As a
nonvanishing tree level term has to be subtracted however, the precision is
limited here. In perturbation theory only the first two universal terms are
known for these couplings. Given these limitations also here reasonable
agreement is found (Fig.~\ref{fig3}, left plot).

{\noindent}\tmtextbf{Acknowledgements}: We thank Peter Weisz for a careful
reading of the manuscript. This work has emerged from the internship project
of M. H. at Humboldt University and we thank the Ecole Normale Superieure
(Paris) for financial support. U.W. acknowledges support by the DFG via SFB
transregio 9.

\appendix\section{Proof of eq. (\ref{Aizlemma})\label{appa}}

We freeze $k \left( l \right) + k' \left( l \right) = K \left( l \right)$ to
fixed values and then show for an arbitrary charge distribution $p (x)$ the
counting identity
\begin{equation}
\sum_{k \leqslant K} \delta_{\partial (K - k), p} \delta_{\partial k, q_{x
   y}}  \prod_l \binom{K (l)}{k (l)} =
  \mathcal{X} (x, y ; K) \sum_{k \leqslant K} \delta_{\partial (K - k), p +
  q_{x y}} \delta_{\partial k, 0}  \prod_l \binom{K (l)}{k (l)} \label{count}
\end{equation}
where the large brackets are binomial coefficients. The sums here run over
values from $0$ to $K (l)$ independently for each $k \left( l \right)$. If $x$
and $y$ are {\tmem{not}} in the same percolation cluster made from bonds with
$K (l) > 0$, then $\mathcal{X}$ vanishes and the same is true for
$\delta_{\partial k, q_{x y}}$ for all $k \leqslant K$, and thus both sides
are zero. We thus only have to consider the case $\mathcal{X}= 1$. Following
Aizenman {\cite{Aizenman:1982ze}} (Lemma 3.2) we momentarily think of a graph
$\mathcal{L}$ of $K (l)$ {\tmem{distinguishable}} lines drawn `over' each link
with (link-wise) cardinality $|\mathcal{L}| = K$. Then on both sides we
{\tmem{count}} the number of distinct subsets $\mathcal{L}' \subseteq
\mathcal{L}$ which are such that the cardinalities $k = |\mathcal{L}' |$ and
$K - k = |\mathcal{L} \backslash \mathcal{L}' |$ obey certain divergence
constraints. The identity is proven now by constructing a one-to-one mapping
between the respective subsets contributing on the right hand side and the
left hand side.

Because of $\mathcal{X}= 1$ there is some chain of specific lines
$\mathcal{L}_{x y}$ connecting $x$ and $y$. If we now have some $\mathcal{L}'$
that contributes to the left hand side, then we map
\begin{equation}
  \mathcal{L}' \rightarrow \mathcal{L}'' =\mathcal{L}' \Delta \mathcal{L}_{x
  y} \equiv \mathcal{L}' \cup \mathcal{L}_{x y} \backslash \{\mathcal{L}' \cap
  \mathcal{L}_{x y} \} .
\end{equation}
One may now show that
\begin{itemize}
  \item if $|\mathcal{L}' |$ satisfies the constraints of the left hand side
  of (\ref{count}) then $|\mathcal{L}'' |$ fulfills those on the right hand
  side,
  
  \item under the same mapping $\mathcal{L}''$ maps back to $\mathcal{L}'$.
\end{itemize}
Therefore (\ref{count}) is now established.

We now multiply both sides of (\ref{count}) with $F [K] w [K]$ where $F$ is
arbitrary and $w$ is defined in (\ref{wdef}). Then we sum over $K$ and change
variables to $k' = K - k$ and $k$ to derive
\begin{equation}
  \sum_{k, k'} F [k + k'] w [k] w [k'] \delta_{\partial k, p}
   \delta_{\partial k', q_{x y}} =
   \sum_{k, k'} \mathcal{X}(x, y ; k + k') F [k + k'] w [k] w [k']
  \delta_{\partial k, p + q_{x y}} \delta_{\partial k', 0} .
\end{equation}
This may be used repeatedly to derive
\[ \sum_{k, k'} F [k + k'] w [k] w [k'] \left\{ \delta_{\partial k, q_{1234}}
   \delta_{\partial k', 0} - \delta_{\partial k, q_{12}} \delta_{\partial k',
   q_{34}} - \delta_{\partial k, q_{13}} \delta_{\partial k', q_{24}} -
   \delta_{\partial k, q_{14}} \delta_{\partial k', q_{23}} \right\} = \]
\begin{equation}
  \sum_{k, k'} F [k + k'] w [k] w [k'] \delta_{\partial k, q_{1234}}
  \delta_{\partial k', 0} \{1 -\mathcal{X}_{34} -\mathcal{X}_{24}
  -\mathcal{X}_{23} \} \assign E
\end{equation}
with the short hand notation $\mathcal{X}_{i j} \equiv \mathcal{X} (x^{\left(
i \right)}, x^{\left( j \right)} ; k + k')$. To satisfy $\partial k =
q_{1234}$ and $\partial k' = 0$, the four points must either belong to one or
to two percolation clusters of $k (l) + k' (l) > 0$. In the latter case it is
easy to verify that $\{1 -\mathcal{X}_{34} -\mathcal{X}_{24} -\mathcal{X}_{23}
\}$ vanishes while in the first case it equals $- 2$. This may now be changed
back to
\begin{equation}
  E = - 2 \sum_{k, k'} F [k + k'] w [k] w [k'] \delta_{\partial k, q_{12}}
  \delta_{\partial k', q_{34}} \mathcal{X}_{13} .
\end{equation}

\section{Perturbative expansion\label{appPT}}

\subsection{Couplings based on correlations\label{appB1}}

We here report details on the computation of the expansion (in $g_0$)
coefficients{\footnote{We count on our readers to not confuse these $p$ with
momenta.}} $p_1 (s, t, z, L)$ and $p_2 (s, t, z, L)$ for the schemes using
$g_s, z_t$ with $s, t \in \{p, a, A\}$ defined in section \ref{betasst}. In
terms of 1PI vertex functions our renormalized parameters read
\begin{equation}
  Y_s^{- 1} = \frac{\Gamma^{(2)}_s (p_s, - p_s) - \Gamma^{(2)}_s (p_s', -
  p_s')}{\hat{p}_s'^2 - \hat{p}_s^2}, \label{Zdef}
\end{equation}
\begin{equation}
  m_t^2 = - Y_t \Gamma^{(2)}_t (p_t, - p_t) - \hat{p}_t^2,
\end{equation}
\begin{equation}
  g_s = - Y_s^2 \Gamma^{(4)}_s (p_s, p_s, - p_s, - p_s),
\end{equation}
with $Y_s$ being the `wave function' renormalization factor and all quantities
are at finite $L$. Standard bare perturbation theory gives (dropping
subscripts $s, t$ here)
\begin{equation}
  \Gamma^{(2)} (p, - p) = - m_0^2 - \hat{p}^2 - g_0 \frac{1}{2} J_1 + g_0^2
  \frac{1}{4} J_1 H_1 (0) + g_0^2 \frac{1}{6} J_2 (p) + \Omicron (g_0^3),
\end{equation}
\begin{equation}
   \Gamma^{(4)} (p, p, - p, - p) = - g_0 + g_0^2  \frac{3}{2} H_1 (p)
  - g_0^3 \left( 3 H_{2, 1} (p) + \frac{3}{4} H_{2, 2} (p) + \frac{3}{2} H_{2,
  3} (p) \right) + \Omicron (g_0^4) .
\end{equation}
The capital letters stand for the usual Feynman diagrams for the two and four
point functions up to two loops and are given explicitly below. The mass
parameter in all propagators is the bare mass $m_0^2$ at this stage. The
actual evaluation proceeds via the following sequence of steps,
\begin{equation}
  \tilde{G} (p) = \frac{1}{\hat{p}^2 + m_0^2}, \hspace{1em} \hat{p}_{\mu} = 2
  \sin (p_{\mu} / 2),
\end{equation}
\begin{equation}
  G_s (x) = \frac{1}{L^4} \sum_{q \in \mathcal{B}_s} \mathe^{i q x} \tilde{G}
  (q),
\end{equation}
\begin{equation}
  J_{1 s} = G_s (0) .
\end{equation}
Note that in
\begin{equation}
  \widetilde{G_s^n}_{} (p) = \sum_x [G_s (x)]^n \mathe^{- i p x}, \hspace{1em}
  n = 2, 3, \ldots \label{powtilde}
\end{equation}
only integer momenta (e.g. $2 p$) are appropriate for even $n$ where $[G_s
(x)]^n$ is periodic for all $s$. We form (dropping subscripts $s$ again)

\begin{equation}
  H_1 (p) = \frac{1}{3} [2 \widetilde{G^2} (0) + \widetilde{G^2} (2 p)]
\end{equation}
and similarly
\begin{equation}
  J_2 (p) = \widetilde{G^3} (p),
\end{equation}
\begin{equation}
  H_{2, 1} (p) = \frac{1}{3 L^4} \sum_{q \in \mathcal{B}_s} \tilde{G} (q)
  \widetilde{G^2} (q - p) [2 \tilde{G} (q) + \tilde{G} (q - 2 p)],
\end{equation}
\begin{equation}
  H_{2, 2} (p) = \frac{1}{3} [2 ( \widetilde{G^2} (0))^2 + ( \widetilde{G^2}
  (2 p))^2],
\end{equation}
\begin{equation}
  H_{2, 3} (p) = J_1 \frac{1}{3 L^4} \sum_q \tilde{G} (q)^2 [2 \tilde{G} (q) +
  \tilde{G} (q - 2 p)] .
\end{equation}
The Fourier transformations are performed as FFT on one coordinate direction
after another and the whole 2 loop computation again has computational
complexity $D L^4 \ln L$ only, see {\cite{Weisz:2010xx}} for more details.

With these expressions we can write (omitting the remainders$\ldots . +
\Omicron (g_0^3)$)
\begin{equation}
  Y_s = 1 + \frac{g_0^2}{6}  \frac{J_{2 s} (p_s') - J_{2 s}
  (p_s)}{\hat{p}_s'^2 - \hat{p}_s^2},
\end{equation}
\begin{equation}
\Delta m_t^2 = m_0^2 - m_t^2 = - \frac{g_0}{2} J_{1 t} + \frac{g_0^2}{4}
   J_{1 t} H_{1 t} +
  \frac{g_0^2}{6} \left[ J_{2 t} (p_t) - \frac{m_0^2 +
  \hat{p}_t^2}{\hat{p}_t'^2 - \hat{p}_t^2} (J_{2 t} (p_t') - J_{2 t} (p_t))
  \right] \label{Deltam2}
\end{equation}
and
\begin{eqnarray}
   g_s &=& g_0 - g_0^2  \frac{3}{2} H_{1 s} (p_s) + g_0^3 \left[ 3 H_{2, 1 s}
   (p_s) + \frac{3}{4} H_{2, 2 s} (p_s) + \right.
   \nonumber \\
  && \left. \frac{3}{2} H_{2, 3 s} (p_s) + \frac{1}{3}  \frac{J_{2 s} (p_s') -
  J_{2 s} (p_s)}{\hat{p}_s'^2 - \hat{p}_s^2} \right] .
\end{eqnarray}
In order to obtain $g_s$ as a function of $g_0$ and $z_t$ we have to combine
now the last two equations to eliminate $m_0^2$ on the right hand sides. To
the order considered and using
\begin{equation}
  \frac{d J_{1 t}}{d m_0^2} = - H_{1 t} (0), \hspace{1em} J_{1 s} \frac{d H_{1
  s} (p)}{d m_0^2} = - 2 H_{2, 3 s} (p) \label{H1der}
\end{equation}
we arrive at
\begin{equation}
  \Delta m^2_t = q_1 (t, z_t, L) g_0 + q_2 (t, z_t, L) g_0^2 \label{Dm2exp}
\end{equation}
with
\begin{eqnarray}
  q_1 (t, z_t, L) & = & - \frac{1}{2} J_{1 t} (m_t^2), \\
  q_2 (t, z_t, L) & = & \frac{1}{6}  \left[ J_{2 t} (m^2_t, p_t) - \frac{m_t^2
  + \hat{p}_t^2}{\hat{p}_t'^2 - \hat{p}_t^2} (J_{2 t} (m^2_t, p_t') - J_{2 t}
  (m^2_t, p_t)) \right] 
\end{eqnarray}
and then at
\begin{equation}
  g = g_0 + p_1 (s, t, z_t, L) g_0^2 + p_2 (s, t, z_t, L) g_0^3
\end{equation}
with
\begin{eqnarray}
  p_1 (s, t, z_t, L) & = & - \frac{3}{2} H_{1 s} (m^2_t, p_s), \\
  p_2 (s, t, z_t, L) & = & 3 H_{2, 1 s} (m^2_t, p_s) + \frac{3}{4} H_{2, 2 s}
  (m^2_t, p_s) + \frac{3}{2} H_{2, 3 s} (m^2_t, p_s) \left[ 1 - \frac{J_{1
  t}}{J_{1 s}} \right] + \nonumber\\
  &  & \frac{1}{3} \frac{J_{2 s} (m^2_t, p_{s'}) - J_{2 s} (m^2_t,
  p_s)}{\hat{p}_{s'}^2 - \hat{p}_s^2} . 
\end{eqnarray}
In these formulae $m_t$ on the right hand sides is given by $z_t / L$, of
course, and $m^2_t$ in the arguments of $J_{1 s}, H_{1
s}, \ldots$ refer to the mass value used here in the propagators that enter.

\subsection{Couplings based on partition function ratios}

We here work out the coefficients appearing in (\ref{ZZPT}). The leading order
is trivially given by
\begin{equation} f_0^{s, \tilde{s}} (z_s, L / a) = - \frac{1}{2} \left( \sum_{p \in
   \mathcal{B}_s} - \sum_{p \in \mathcal{B}_{\tilde{s}}} \right) \ln (
   \hat{p}^2 + m_s^2) . 
\end{equation}
The first correction receives contributions from both the interaction term and
from eliminating $m_0^2$ for the renormalized mass $m_s^2$ by (\ref{Deltam2})
and we find after simple steps
\begin{equation}
  f_1^{s, \tilde{s}} (z_s, L / a) = \frac{L^4}{8} [G_s (0) - G_{\tilde{s}}
  (0)]^2
\end{equation}
with the mass $m_s$ here in both propagators. Numerical values are given in
Table~\ref{tabz2},

\begin{table}[htb]
  \begin{center}
  
  \begin{tabular}{|l|l|l|l|l|}
    \hline
    $L$ & \multicolumn{1}{|c|}{$f_0^{a, A}$} & $f_1^{a, A} \times 10^3$ &
    \multicolumn{1}{|c|}{$- f_0^{a, p}$} & $f_1^{a,
    p} \times 10^3$\\
    \hline
    \phantom{1}8 & 0.26076127 & 0.35227920 & 0.30274830 & 2.4978979\\
    \hline
    10 & 0.24731850 & 0.33411586 & 0.29837653 & 2.4880152 \\
    \hline
    12 & 0.24059900 & 0.32490117 & 0.29621943 & 2.4832198\\
    \hline
    16 & 0.23423877 & 0.31612031 & 0.29419783 & 2.4787956\\
    \hline
    22 & 0.23052145 & 0.31096912 & 0.29302625 & 2.4762694\\
    \hline
    32 & 0.22836421 & 0.30797464 & 0.29234996 & 2.4748253\\
    \hline
    64 & 0.22693150 & 0.30598406 & 0.29190233 & 2.4738756\\
    \hline
  \end{tabular}
  \caption{Perturbative coefficients $f^{a, A}_0, f^{a, A}_1$ for $z_a =
  2$.\label{tabz2}}
\end{center}
\end{table}

and asymptotic large $L$ Symanzik expansions for these cases are
\begin{equation}
  f^{a, A}_0 = 0.226457 + 1.941 L^{- 2} + \Omicron (L^{- 4}),
\end{equation}
\begin{equation}
  f^{a, A}_1 \times 10^3 = 0.305324 + 2.698 L^{- 2} + \Omicron (L^{- 4}),
\end{equation}
\begin{equation}
  - f^{a, p}_0 = 0.291754 + 0.6049 L^{- 2} + \Omicron (L^{- 4}),
\end{equation}
\begin{equation}
  f^{a, p}_1 \times 10^3 = 2.47356 + 1.278 L^{- 2} + \Omicron (L^{- 4}) .
\end{equation}

\end{document}